\documentclass[sigconf]{acmart}
\AtBeginDocument{%
  }

\copyrightyear{2025}
\acmYear{2025}
\setcopyright{rightsretained}
\acmConference[SIGIR '25]{Proceedings of the 48th International ACM SIGIR Conference on Research and Development in Information Retrieval}{July 13--18, 2025}{Padua, Italy}
\acmBooktitle{Proceedings of the 48th International ACM SIGIR Conference on Research and Development in Information Retrieval (SIGIR '25), July 13--18, 2025, Padua, Italy}\acmDOI{10.1145/3726302.3730135}
\acmISBN{979-8-4007-1592-1/2025/07}

\begin{document}



\title[Tevatron 2.0: Unified Document Retrieval Toolkit across Scale, Language, and Modality]{Tevatron 2.0: Unified Document Retrieval Toolkit \\
across Scale, Language, and Modality}


\author{Xueguang Ma}\authornote{The first three authors contributed equally.}
\email{x93ma@uwaterloo.ca}
\affiliation{%
  \institution{University of Waterloo}
  \city{Waterloo}
  \country{Canada}
}
\orcid{0000-0003-3430-4910}

\author{Luyu Gao}
\authornotemark[1]
\email{luyug@cs.cmu.edu}
\affiliation{%
  \institution{Carnegie Mellon University}
  \city{Pittsburgh}
  \country{United States}
}

\author{Shengyao Zhuang}
\authornotemark[1]
\email{shengyao.zhuang@csiro.au}
\affiliation{%
  \institution{CSIRO}
  \city{Queensland}
  \country{Australia}
}

\author{Jiaqi Samantha Zhan}
\email{j35zhan@uwaterloo.ca}
\affiliation{%
  \institution{University of Waterloo}
  \city{Waterloo}
  \country{Canada}
}
\author{Jamie Callan}
\email{callan@cs.cmu.edu}
\affiliation{%
  \institution{Carnegie Mellon University}
  \city{Pittsburgh}
  \country{United States}
}
\author{Jimmy Lin}
\email{jimmylin@uwaterloo.ca}
\affiliation{%
  \institution{University of Waterloo}
  \city{Waterloo}
  \country{Canada}
}
\settopmatter{authorsperrow=4}



\begin{abstract}
Recent advancements in large language models (LLMs) have driven interest in billion-scale retrieval models with strong generalization across retrieval tasks and languages. Additionally, progress in large vision-language models has created new opportunities for multimodal retrieval. 
In response, we have updated the Tevatron toolkit, introducing a unified pipeline that enables researchers to explore retriever models at different scales, across multiple languages, and with various modalities.
This demo paper highlights the toolkit’s key features, bridging academia and industry by supporting efficient training, inference, and evaluation of neural retrievers. We showcase a unified dense retriever achieving strong multilingual and multimodal effectiveness, and conduct a cross-modality zero-shot study to demonstrate its research potential.
Alongside, we release OmniEmbed, to the best of our knowledge, the first embedding model that unifies text, image document, video, and audio retrieval, serving as a baseline for future research.
\end{abstract}



\maketitle

\section{Introduction}
The Tevatron toolkit~\cite{tevatronv1} is initially developed in 2021 during the BERT era, a period marked by the rise of neural dense retrievers, represented by models like Dense Passage Retrieval (DPR)~\cite{karpukhin-etal-2020-dense}, which became a focal point in information retrieval research.
Tevatron is designed to provide an efficient and flexible framework, enabling researchers to easily modify or integrate new model architectures and training strategies for their specific needs.
Since the initial release, dozens of previously published works have cited its usage.
These works span a wide range of topics, including diverse text representation methods, multilingual and cross-lingual retrieval, and retrieval-augmented generation (RAG) pipelines.

Recent advancements in large language models (LLMs) have introduced new opportunities and challenges in the field of information retrieval. Billion-scale retrieval models have demonstrated superior generalizability across tasks and languages compared to their BERT-era counterparts~\cite{repllama,wang2024improvingtextembeddingslarge}. Additionally, the emergence of multimodal large language models~\cite{qwen2.5-VL,Qwen2.5-Omni} has opened new possibilities for multimodal retrieval, leveraging large vision-language models to eliminate the need for document parsing and the cumbersome maintenance of disparate modalities~\cite{ma-etal-2024-unifying,faysse2024colpaliefficientdocumentretrieval,lin2025universal}. 
These developments point toward a promising direction: the creation of unified retrieval systems capable of handling diverse tasks, languages, and modalities.

However, several challenges hinder the exploration of these advancements. First, training billion-scale retrieval models requires significantly more GPU memory, posing a barrier to researchers with limited computational resources.
Second, while unified retrieval across modalities is an attractive goal, existing data formats designed for text retrieval are often ill-suited for organizing multimodal data. Third, as dense retrievers increasingly serve as industry solutions, there is a growing demand for more efficient inference methods to meet real-world deployment requirements.

In this work, we present updates to the Tevatron toolkit that address these challenges. Our contributions include a new data organization to support unified multimodal retriever training, as well as the integration of best practices from general LLM training and inference to enhance memory and computational efficiency. 
To facilitate further research, we have converted multiple open-source datasets into a unified format and hosted them in the Tevatron Hugging Face collection, which we will maintain and update continuously.
We also demonstrate Tevatron’s capability to support future research by training a unified dense retriever capable of multilingual and multimodal retrieval and providing analysis such as cross-modality generalization.

\section{Tevatron-v2 Overview}
\subsection{Unified Data Management}
Data management is a critical aspect often overlooked in individual research projects. As dense retrieval increasingly aims to handle generalizable tasks, it becomes essential to efficiently explore multiple sets of training data and their combinations.

Tevatron-v1 is specifically designed for text retrieval models, where training data for different text retrieval tasks could be unified into the following format for contrastive learning:

{\small
\begin{verbatim}
{
   "query_id": "<query id>",
   "query": "<query text>",
   "positive_passages": [
     {"docid": "<passage id>", "title": "<passage title>", 
      "text": "<passage body>"}, ...],
   "negative_passages": [
     {"docid": "<passage id>", "title": "<passage title>", 
      "text": "<passage body>"}, ...]
}
\end{verbatim}
}

\noindent In this format, the raw document content for both positive and negative candidates is stored alongside each query instance. While this provides transparency for users to explore and debug the training data, it is not well-suited for multimodality data (where documents or queries can be images) or mixed-modality data (where queries and documents can contain both text and images).

As the number of queries expands, and each query is typically associated with multiple hard negative documents, the document lists across multiple queries often contain duplicate content. For instance, if each document in a corpus is associated with a query, and we aim to build training data with 20 hard negatives per query, the storage requirement for the training data becomes 20 times that of the corpus. This storage overhead becomes particularly prohibitive for image data, which is significantly more costly to store compared to text.

To address these challenges, we introduce a new unified data format for Tevatron-v2:
{\small
\begin{verbatim}
query:
{
   "query_id": "<query id>",
   "query_text": "<query text>",
   "query_image": "<query image>",
   "query_video": "<path to video>",
   "query_audio": "<path to audio>"
   "positive_document_ids": ["<document id>", ...],
   "negative_document_ids": ["<document id>", ...],
}
\end{verbatim}
}

{\small
\begin{verbatim}
corpus:
{
   "docid": "<document id>",
   "document_text": "<document text>",
   "document_image": "<document image>",
   "document_video": "<document video path>",
   "document_audio": "<document audio path>",
}
\end{verbatim}
}
\noindent In this new format, we decouple the training queries from the corpus. For document candidates associated with each query, we only store document IDs, dynamically loading the raw content within the dataloader. This approach significantly reduces storage requirements.

Additionally, this format efficiently organizes various data modality combinations without introducing complexity. It is compatible with text-only retrieval data, image-only retrieval data, and mixed-modality retrieval data.
If a query or document contains more than one piece of text or image, we assume they can be merged into a single representation.

The dataloader and collator in Tevatron-v2 are designed to be modality-agnostic, enabling seamless training across different data combinations. For example, mixing text-only and image retrieval datasets allows training queries to encounter documents from varied modalities, helping to reduce modality bias. To support flexible data integration, Tevatron-v2 also introduces a Multi-Dataset class, allowing users to configure multiple training datasets.

\subsection{GPU Memory Efficiency}

\begin{table}[t]
\centering
\resizebox{0.45\textwidth}{!}{
\begin{tabular}{lllll}
\toprule
Finetune & ZeRO         & FlashAttn & GPU Memory      & Training Time \\
\midrule
Full-FT     & ZeRO0        & No         & OOM             & --            \\
Full-FT     & ZeRO3        & No         & 63,146 MiB $\times$4    & 27 hours      \\
Full-FT     & ZeRO3        & Yes        & 62,916 MiB $\times$4    & 26 hours      \\
Full-FT     & ZeRO3-off    & Yes        & 21,764 MiB $\times$4    & 44 hours      \\
LoRA        & ZeRO0        & No         & 28,554 MiB $\times$4    & 19 hours      \\
LoRA        & ZeRO0        & Yes        & 28,172 MiB $\times$4    & 18 hours      \\
LoRA        & ZeRO3        & Yes        & 25,778 MiB $\times$4    & 25 hours      \\
LoRA        & ZeRO3        & Yes        & 69,324 MiB $\times$1    & 74 hours      \\
\bottomrule
\end{tabular}
}
\caption{GPU memory and training time for fine-tuning Llama3.1-8B as a dense retriever using Tevatron-v2 with Rep-Llama~\cite{repllama} recipe on MS MARCO Passage.}
\label{tab:finetune}
\vspace{-1.1cm}
\end{table}

The primary challenge in training billion-scale dense retrieval models lies in their substantial GPU memory requirements. Additionally, retriever training benefits from larger batch sizes and a large set of hard negative candidates for each update, further escalating these memory demands. Tevatron-v2 integrates several best practices from general large language model (LLM) training, including LoRA (Low-Rank Adaptation)~\cite{hu2022lora}, DeepSpeed ZeRO optimization~\cite{deepspeed}, and FlashAttention~\cite{dao2022flashattention}, to address these challenges.

Table~\ref{tab:finetune} demonstrates that each memory-efficient strategy significantly reduces GPU memory usage.
For example, full fine-tuning without optimization results in out-of-memory errors.
Enabling ZeRO stage 3 reduces memory usage to 63 GB per GPU on a 4-GPU machine, with a training time of 27 hours.
With CPU offloading (zero3-off), memory usage drops to 22 GB per GPU, though training time increases to 44 hours due to offloading overhead.

For LoRA-based fine-tuning, the memory usage is much lower than full fine-tuning. 
Combining LoRA with ZeRO stage 3 and FlashAttention further reduces memory needs.
Notably, training on a single GPU with LoRA, zero3, and FlashAttention is feasible, consuming 69,324 MiB of memory in charge of longer training time, but it makes things possible to explore on very limited compute resources.
FlashAttention generally provides greater benefits in longer-context scenarios. However, in the MS MARCO Passage dataset, where documents are short (averaging 60 words), its impact on speed and memory cost is less pronounced.


\subsection{Inference Efficiency}
\begin{figure}[h]
    \centering
    \includegraphics[width=0.9\linewidth]{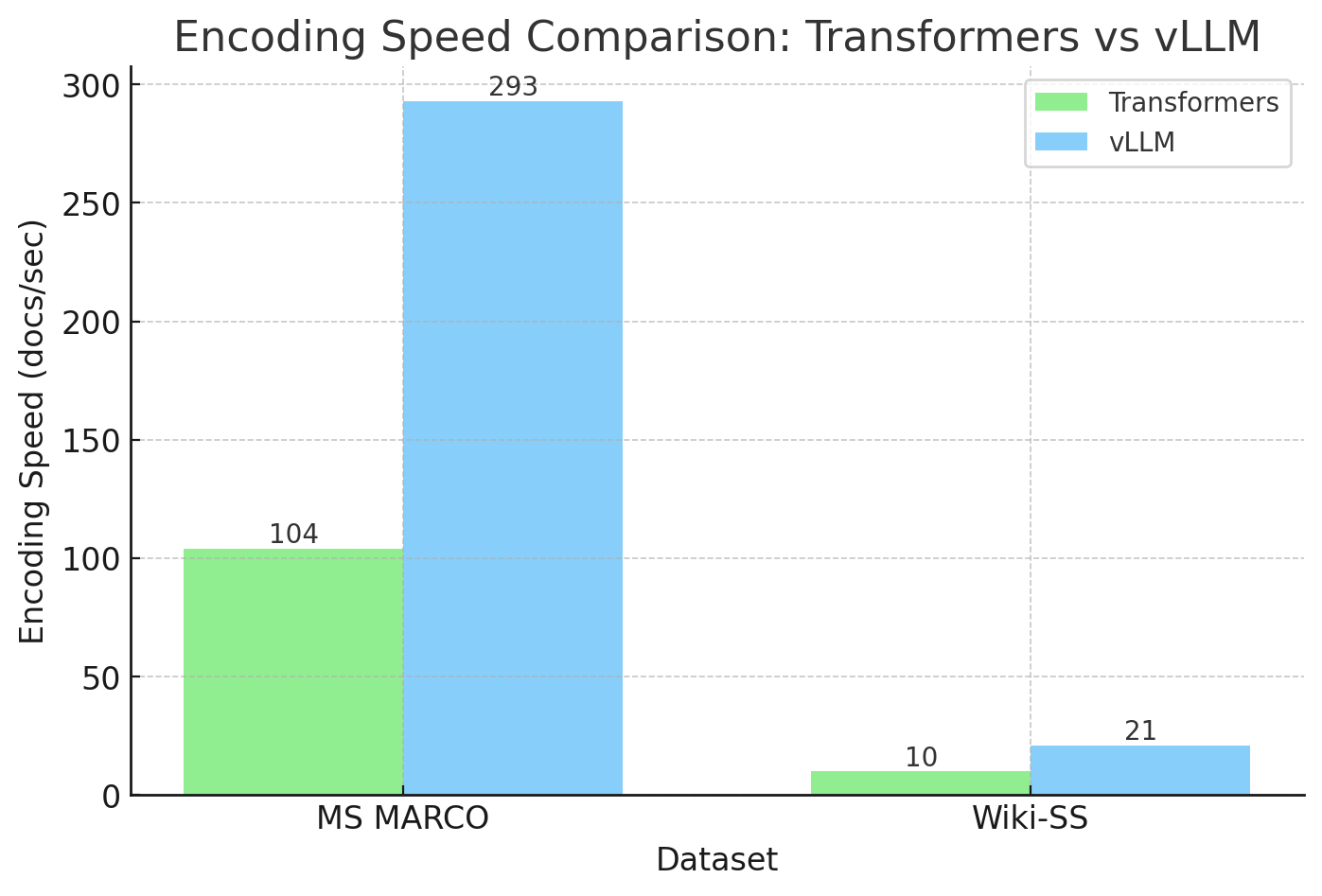}
    \caption{Encoding speed comparison between Transformers and vLLM implementation. For text, we used a retriever based on Llama3.1-8B retriever; for Wiki-SS, we used a multimodal retriever based on QWen2-VL-2B retriever with 784×784 image inputs.}
    \label{fig:encoding_speed}
\end{figure}



vLLM~\cite{vllm} is an optimized serving and inference library for LLMs, designed to deliver high-throughput and low-latency inference. 
Integrating vLLM into Tevatron-v2 offers two key advantages. First, it significantly improves encoding speed, making the process more efficient. Second, it simplifies the deployment of models trained with the Tevatron codebase, enabling users to deploy them more easily and effectively. Additionally, vLLM majorly supports generation models, allowing retrievers to work closely with generation models in frameworks like retrieval augmented generation. This integration makes such collaborative workflows more seamless and efficient.

To quantitatively evaluate the benefits of vLLM, we compared the encoding speed of vLLM and the standard Transformers implementation in two settings: text encoding on the MS MARCO passage ranking corpus and document image encoding on the Wiki-SS (Wikipedia screenshot corpus). For text encoding, we used an 8B LLM retriever based on the Llama3.1-8B architecture. For Wiki-SS encoding, we employed a multimodal retriever based on the QWen2-VL-2B backbone, with images input at a resolution of 784x784.
As shown in Figure~\ref{fig:encoding_speed}, vLLM outperforms the Transformers implementation in both text and image encoding tasks by around 3 times of encoding speed.

Besides encoding latency, as the size of hidden states in large language models (LLMs) continues to grow, the increasing dimensionality of text embeddings raises concerns about storage costs for corpus indexes and search latency. 
To address this challenge, we integrate Matryoshka Representation Learning (MRL)~\cite{kusupati2022matryoshka} into Tevatron-v2. This approach enables the model to learn nested, scalable representations, where embeddings can be truncated to smaller dimensionalities without significant effectiveness degradation. During training, users can specify the target dimensionalities they wish to optimize for, allowing the model to adapt to diverse requirements. At inference time, this flexibility empowers users to dynamically adjust the text embedding dimensionality based on specific storage or computational constraints.

\subsection{Comparison with Other Toolkits}

The IR community has developed a number of outstanding toolkits that have greatly advanced research and experimentation. For example, Pyserini enables reproducible information retrieval experiments with robust integration of sparse and dense retrievers~\cite{pyserini}. MTEB provides a comprehensive benchmark suite for evaluating embedding models across a wide range of retrieval and general embedding tasks~\cite{muennighoff-etal-2023-mteb}. Toolkits like sentence-transformers~\cite{reimers-2019-sentence-bert}, BGE~\cite{chen-etal-2024-m3} and RAG-Retrieval~\cite{stella} offer flexible and performant implementations of state-of-the-art retrievers and rerankers, making them widely adopted in both academia and industry.
There are also specialized toolkits designed for LLM-based reranking~\cite{rankllm, sun-etal-2023-chatgpt, zhang2025rankwogpt}.

Tevatron aims to offer an extensible framework that supports rapid prototyping and scalable training, with modular components for data processing, model training, and evaluation. It is designed with flexibility in mind, making it easy to incorporate new training paradigms, languages, and modalities. This enables researchers to explore new scenarios within a single, cohesive framework, paving the way for more generalizable and powerful retrieval systems.

\section{Experiment: Unified Dense Retriever}
To demonstrate the capabilities of our toolkit, we present the training of a dense retrieval model using a set of Tevatron self-contained data (available on Hugging Face) to develop a unified multimodal and multilingual dense retriever.
We further explore in-modality zero-shot and cross-modality zero-shot settings to highlight the potential of Tevatron-v2 for future research.

\subsection{Setup}
\subsubsection{Training data}
\begin{itemize}
    \item BGE-Training Data~\cite{chen-etal-2024-m3}: A set of text retrieval training data, such as web search (MS MARCO Passage), Wikipedia-based QA and Fact verification (NQ, FEVER), scientific document retrieval (SCIDOCS), etc. 1.84 million in total derived from bge-full-training data.
    \item WikiSS~\cite{ma-etal-2024-unifying}: This dataset, released as part of the DSE work, is based on natural question query sets and the Wikipedia screenshot corpus. It includes 29.3k training samples.
    \item PixMo-docs:  PixMo-Docs~\cite{deitke2024molmopixmoopenweights} is originally a collection of synthetic question-answer pairs about computer-generated images (e.g., charts, tables, diagrams, and documents), we converted PixMo-Docs into retrieval training data through filtering and hard negative mining.
We removed questions that were too specific and unsuitable for open-domain question answering (e.g., ``What is the summary of the text?''). This was done by leveraging DSE as a retriever to search for each query. If the corresponding paired document did not appear in the top-100 retrieval results, the query was deemed unsuitable and removed.
We identified challenging negative examples for each question by randomly sampling non-positive documents from the top-100 retrieval results.
After processing, we obtained 1.75 million training samples.
    \item ColPali-training-data~\cite{faysse2024colpaliefficientdocumentretrieval}: This dataset is the training data for the ColPali model, a multi-vector document screenshot retriever. It includes 127k query-image pairs. The original release did not contain hard negatives, so we followed a similar approach as above to mine hard negative candidates.
    \item MSR-VTT~\cite{msrvtt} and AudioCaps~\cite{audiocaps} are video and audio retrieval datasets, respectively. We include them to support training the Omni variants of our model. See additional details below.
\end{itemize}

\subsubsection{Hyperparameters}

The Tevatron-VL model, based on Qwen2.5-VL-3B-Instruct~\cite{qwen2.5-VL}, is trained on the combined text and visual datasets for one epoch. The Tevatron-Omni model, based on Qwen2.5-Omni-7B~\cite{Qwen2.5-Omni}, additionally incorporates video and audio data.
Training is performed on a machine with 8×NVIDIA H100 GPUs using a batch size of 128 queries. Each query is paired with one positive and three hard negative documents. We adopt LoRA for memory-efficient training.

\subsubsection{Evaluation}
To assess the model's effectiveness on retrieval tasks across languages and modalities, we evaluate it on the following widely used retrieval evaluation datasets:

\begin{itemize}
\item BEIR~\cite{thakur2021beir}: A diverse collection of English retrieval tasks. We evaluate on its 13 publicly available datasets, with effectiveness reported in nDCG@10.
\item MIRACL~\cite{MIRACL}: A multilingual retrieval evaluation benchmark covering 18 languages, with effectiveness reported in nDCG@10.
\item ViDoRe~\cite{faysse2024colpaliefficientdocumentretrieval}: A multimodal evaluation benchmark that includes table, chart, and document screenshot retrieval tasks. Effectiveness is reported in nDCG@5.
\end{itemize}

\noindent Note that Tevatron-v2 also self-contains evaluation scripts for above evaluation.

\subsubsection{Model Variants}
To further investigate the effectiveness of in-modality and cross-modality zero-shot generalization, we trained two additional variants:

\begin{itemize}
\item Tevatron-WikiSS, which is trained exclusively on Wikipedia screenshot-based document image retrieval data. This variant is evaluated using the ViDoRe benchmark for in-modality zero-shot effectiveness.
\item Tevatron-BGE, which is trained solely on text retrieval data to adapt the multimodal backbone as a retriever. This variant is evaluated using the ViDoRe benchmark for cross-modality zero-shot effectiveness.
\end{itemize}

\subsection{Results}

\begin{table}[t]
\centering
\resizebox{1\columnwidth}{!}{%
\begin{tabular}{ l | c | c | c | c}
\toprule
 Method & Base Model & BEIR  & ViDoRe & MIRACL\\ \midrule
 BGE-M3~\cite{chen-etal-2024-m3} & X-RoBERTa & 50.0 & 66.1$^*$ & 69.2 \\  
 MistralE5~\cite{wang2024improvingtextembeddingslarge} & Mistral0.1-7B & 59.0 & - & 62.2 \\
 DSE-QWen2~\cite{ma-etal-2024-unifying} & Qwen2vl-2B & - & 85.8 & - \\ 
 GME-2B~\cite{gme} & Qwen2vl-2B & 55.4 & 87.8 & - \\ 
 \midrule
 Tevatron-WikiSS & Qwen2.5vl-3B & 40.8 & 73.3 & 30.6 \\  
 Tevatron-BGE & Qwen2.5vl-3B & 57.0 & 76.4  & 67.5 \\
 Tevatron-VL & Qwen2.5vl-3B & 54.3 & 85.3  & 66.5 \\
Tevatron-Omni & Qwen2.5Omni & 58.2 & 85.8  & 69.1 \\
 
\bottomrule
\end{tabular}
}
\caption{Multimodal and multilingual retrieval results. $^*$The result of BGE-M3 on ViDoRe is based on OCR.}
\label{tab:results}
\vspace{-0.7cm}
\end{table}

\begin{table}[h]
\centering
\resizebox{0.7\columnwidth}{!}{%
\begin{tabular}{ l | c | c }
\toprule
 Method & MSR-VTT & AudioCaps \\
\midrule
CLIP~\cite{clip4video} & 31.2 & - \\
CE~\cite{CE} & - & 23.1 \\
Tevatron-Omni & 51.3 & 34.0 \\
\bottomrule
\end{tabular}
}
\caption{Recall@1 of video and audio retrieval tasks of Tevatron-Omni compared to baselines.}
\label{tab:results-video}
\vspace{-1cm}
\end{table}

Table~\ref{tab:results} presents a comparison of retrieval effectiveness across various models, highlighting the effectiveness of our unified retriever model, Tevatron-VL and Tevatron-Omni, against other existing representative models.
The results demonstrate that our unified retriever achieves competitive effectiveness across multiple dimensions of retrieval tasks, including English multi-task retrieval, multilingual retrieval, and multimodal retrieval.
In addition, our Tevatron-Omni also effectively supports video and audio embedding as illustrated in Table~\ref{tab:results-video}.
Existing retrieval models are typically optimized for one or two specific dimensions. 
For instance, BGE-M3~\cite{chen-etal-2024-m3} and MistralE5~\cite{wang2024improvingtextembeddingslarge} are primarily focused on text retrieval tasks.
In contrast, Tevatron is designed to easily handle all these optimizations across retrieval tasks, languages, and modalities.
Besides, our training data and training pipeline are fully open-sourced.

\paragraph{Zero-Shot Generalization}
An exciting finding from the Tevatron-BGE model variants, which trains a large vision-language model backbone using only text retrieval data, demonstrates strong cross-modal zero-shot effectiveness on ViDoRe. The effectiveness score is even higher than that of in-modality zero-shot when using WikiSS as training data. Since the BGE data encompasses a more diverse range of retrieval tasks, this suggests that the backbone model excels in aligning textual and visual inputs. Furthermore, it highlights that fine-tuning the model to learn relevancy is more critical than focusing on modality alignment. This implies that even in the absence of multimodal retrieval training data, training on diverse text-only retrieval tasks can potentially yield effective multimodal retrieval. This approach may prove more effective than relying on text-based retrieval combined with OCR, as seen in models like BGE-M3 on ViDoRe.
This showcases the flexibility and research-friendly design of Tevatron-v2, empowering researchers to experiment with diverse training data and tasks while exploring new research questions.

\section{Conclusion}
In this paper, we introduce Tevatron-v2, a unified and efficient toolkit for advancing large-scale, multimodal, and multilingual retrieval research. 
By demonstrating the training of a unified dense retriever capable of handling both multilingual text and image document retrieval, Tevatron-v2 showcases its versatility and potential to drive future research in retrieval, making it a valuable tool for both academia and industry.
Tevatron-v2 is fully open-sourced at \url{https://github.com/texttron/tevatron}.

\section*{Acknowledgments}
We sincerely thank the open-source community, as well as the contributors and users who have supported the development. This research was supported in part by the Natural Sciences and Engineering Research Council (NSERC) of Canada.


\balance
\bibliographystyle{ACM-Reference-Format}
\bibliography{citation}


\begin{thebibliography}{30}


\ifx \showCODEN    \undefined \def \showCODEN     #1{\unskip}     \fi
\ifx \showISBNx    \undefined \def \showISBNx     #1{\unskip}     \fi
\ifx \showISBNxiii \undefined \def \showISBNxiii  #1{\unskip}     \fi
\ifx \showISSN     \undefined \def \showISSN      #1{\unskip}     \fi
\ifx \showLCCN     \undefined \def \showLCCN      #1{\unskip}     \fi
\ifx \shownote     \undefined \def \shownote      #1{#1}          \fi
\ifx \showarticletitle \undefined \def \showarticletitle #1{#1}   \fi
\ifx \showURL      \undefined \def \showURL       {\relax}        \fi
\providecommand\bibfield[2]{#2}
\providecommand\bibinfo[2]{#2}
\providecommand\natexlab[1]{#1}
\providecommand\showeprint[2][]{arXiv:#2}

\bibitem[Chen et~al\mbox{.}(2024)]%
        {chen-etal-2024-m3}
\bibfield{author}{\bibinfo{person}{Jianlyu Chen}, \bibinfo{person}{Shitao Xiao}, \bibinfo{person}{Peitian Zhang}, \bibinfo{person}{Kun Luo}, \bibinfo{person}{Defu Lian}, {and} \bibinfo{person}{Zheng Liu}.} \bibinfo{year}{2024}\natexlab{}.
\newblock \showarticletitle{{M}3-Embedding: Multi-Linguality, Multi-Functionality, Multi-Granularity Text Embeddings Through Self-Knowledge Distillation}. In \bibinfo{booktitle}{\emph{Findings of the Association for Computational Linguistics: ACL 2024}}, \bibfield{editor}{\bibinfo{person}{Lun-Wei Ku}, \bibinfo{person}{Andre Martins}, {and} \bibinfo{person}{Vivek Srikumar}} (Eds.). \bibinfo{publisher}{Association for Computational Linguistics}, \bibinfo{address}{Bangkok, Thailand}, \bibinfo{pages}{2318--2335}.
\newblock


\bibitem[Dao et~al\mbox{.}(2022)]%
        {dao2022flashattention}
\bibfield{author}{\bibinfo{person}{Tri Dao}, \bibinfo{person}{Daniel~Y Fu}, \bibinfo{person}{Stefano Ermon}, \bibinfo{person}{Atri Rudra}, {and} \bibinfo{person}{Christopher Re}.} \bibinfo{year}{2022}\natexlab{}.
\newblock \showarticletitle{FlashAttention: Fast and Memory-Efficient Exact Attention with {IO}-Awareness}. In \bibinfo{booktitle}{\emph{Advances in Neural Information Processing Systems}}, \bibfield{editor}{\bibinfo{person}{Alice~H. Oh}, \bibinfo{person}{Alekh Agarwal}, \bibinfo{person}{Danielle Belgrave}, {and} \bibinfo{person}{Kyunghyun Cho}} (Eds.).
\newblock


\bibitem[Deitke et~al\mbox{.}(2024)]%
        {deitke2024molmopixmoopenweights}
\bibfield{author}{\bibinfo{person}{Matt Deitke}, \bibinfo{person}{Christopher Clark}, \bibinfo{person}{Sangho Lee}, \bibinfo{person}{Rohun Tripathi}, \bibinfo{person}{Yue Yang}, \bibinfo{person}{Jae~Sung Park}, \bibinfo{person}{Mohammadreza Salehi}, \bibinfo{person}{Niklas Muennighoff}, \bibinfo{person}{Kyle Lo}, \bibinfo{person}{Luca Soldaini}, \bibinfo{person}{Jiasen Lu}, \bibinfo{person}{Taira Anderson}, \bibinfo{person}{Erin Bransom}, \bibinfo{person}{Kiana Ehsani}, \bibinfo{person}{Huong Ngo}, \bibinfo{person}{YenSung Chen}, \bibinfo{person}{Ajay Patel}, \bibinfo{person}{Mark Yatskar}, \bibinfo{person}{Chris Callison-Burch}, \bibinfo{person}{Andrew Head}, \bibinfo{person}{Rose Hendrix}, \bibinfo{person}{Favyen Bastani}, \bibinfo{person}{Eli VanderBilt}, \bibinfo{person}{Nathan Lambert}, \bibinfo{person}{Yvonne Chou}, \bibinfo{person}{Arnavi Chheda}, \bibinfo{person}{Jenna Sparks}, \bibinfo{person}{Sam Skjonsberg}, \bibinfo{person}{Michael Schmitz}, \bibinfo{person}{Aaron Sarnat}, \bibinfo{person}{Byron
  Bischoff}, \bibinfo{person}{Pete Walsh}, \bibinfo{person}{Chris Newell}, \bibinfo{person}{Piper Wolters}, \bibinfo{person}{Tanmay Gupta}, \bibinfo{person}{Kuo-Hao Zeng}, \bibinfo{person}{Jon Borchardt}, \bibinfo{person}{Dirk Groeneveld}, \bibinfo{person}{Crystal Nam}, \bibinfo{person}{Sophie Lebrecht}, \bibinfo{person}{Caitlin Wittlif}, \bibinfo{person}{Carissa Schoenick}, \bibinfo{person}{Oscar Michel}, \bibinfo{person}{Ranjay Krishna}, \bibinfo{person}{Luca Weihs}, \bibinfo{person}{Noah~A. Smith}, \bibinfo{person}{Hannaneh Hajishirzi}, \bibinfo{person}{Ross Girshick}, \bibinfo{person}{Ali Farhadi}, {and} \bibinfo{person}{Aniruddha Kembhavi}.} \bibinfo{year}{2024}\natexlab{}.
\newblock \showarticletitle{Molmo and PixMo: Open Weights and Open Data for State-of-the-Art Vision-Language Models}.
\newblock \bibinfo{journal}{\emph{arXiv:2409.17146}} (\bibinfo{year}{2024}).
\newblock


\bibitem[Faysse et~al\mbox{.}(2024)]%
        {faysse2024colpaliefficientdocumentretrieval}
\bibfield{author}{\bibinfo{person}{Manuel Faysse}, \bibinfo{person}{Hugues Sibille}, \bibinfo{person}{Tony Wu}, \bibinfo{person}{Bilel Omrani}, \bibinfo{person}{Gautier Viaud}, \bibinfo{person}{Céline Hudelot}, {and} \bibinfo{person}{Pierre Colombo}.} \bibinfo{year}{2024}\natexlab{}.
\newblock \showarticletitle{ColPali: Efficient Document Retrieval with Vision Language Models}.
\newblock \bibinfo{journal}{\emph{arXiv:2407.01449}} (\bibinfo{year}{2024}).
\newblock


\bibitem[Gao et~al\mbox{.}(2023)]%
        {tevatronv1}
\bibfield{author}{\bibinfo{person}{Luyu Gao}, \bibinfo{person}{Xueguang Ma}, \bibinfo{person}{Jimmy Lin}, {and} \bibinfo{person}{Jamie Callan}.} \bibinfo{year}{2023}\natexlab{}.
\newblock \showarticletitle{Tevatron: An Efficient and Flexible Toolkit for Neural Retrieval}. In \bibinfo{booktitle}{\emph{Proceedings of the 46th International ACM SIGIR Conference on Research and Development in Information Retrieval}} (Taipei, Taiwan) \emph{(\bibinfo{series}{SIGIR '23})}. \bibinfo{publisher}{Association for Computing Machinery}, \bibinfo{address}{New York, NY, USA}, \bibinfo{pages}{3120–3124}.
\newblock
\showISBNx{9781450394086}


\bibitem[Hu et~al\mbox{.}(2022)]%
        {hu2022lora}
\bibfield{author}{\bibinfo{person}{Edward~J. Hu}, \bibinfo{person}{Yelong Shen}, \bibinfo{person}{Phillip Wallis}, \bibinfo{person}{Zeyuan Allen-Zhu}, \bibinfo{person}{Yuanzhi Li}, \bibinfo{person}{Shean Wang}, \bibinfo{person}{Lu Wang}, {and} \bibinfo{person}{Weizhu Chen}.} \bibinfo{year}{2022}\natexlab{}.
\newblock \showarticletitle{Lo{RA}: Low-Rank Adaptation of Large Language Models}. In \bibinfo{booktitle}{\emph{International Conference on Learning Representations}}.
\newblock


\bibitem[Karpukhin et~al\mbox{.}(2020)]%
        {karpukhin-etal-2020-dense}
\bibfield{author}{\bibinfo{person}{Vladimir Karpukhin}, \bibinfo{person}{Barlas Oguz}, \bibinfo{person}{Sewon Min}, \bibinfo{person}{Patrick Lewis}, \bibinfo{person}{Ledell Wu}, \bibinfo{person}{Sergey Edunov}, \bibinfo{person}{Danqi Chen}, {and} \bibinfo{person}{Wen-tau Yih}.} \bibinfo{year}{2020}\natexlab{}.
\newblock \showarticletitle{Dense Passage Retrieval for Open-Domain Question Answering}. In \bibinfo{booktitle}{\emph{Proceedings of the 2020 Conference on Empirical Methods in Natural Language Processing (EMNLP)}}, \bibfield{editor}{\bibinfo{person}{Bonnie Webber}, \bibinfo{person}{Trevor Cohn}, \bibinfo{person}{Yulan He}, {and} \bibinfo{person}{Yang Liu}} (Eds.). \bibinfo{publisher}{Association for Computational Linguistics}, \bibinfo{address}{Online}, \bibinfo{pages}{6769--6781}.
\newblock


\bibitem[Kim et~al\mbox{.}(2019)]%
        {audiocaps}
\bibfield{author}{\bibinfo{person}{Chris~Dongjoo Kim}, \bibinfo{person}{Byeongchang Kim}, \bibinfo{person}{Hyunmin Lee}, {and} \bibinfo{person}{Gunhee Kim}.} \bibinfo{year}{2019}\natexlab{}.
\newblock \showarticletitle{{A}udio{C}aps: Generating Captions for Audios in The Wild}. In \bibinfo{booktitle}{\emph{Proceedings of the 2019 Conference of the North {A}merican Chapter of the Association for Computational Linguistics: Human Language Technologies, Volume 1 (Long and Short Papers)}}, \bibfield{editor}{\bibinfo{person}{Jill Burstein}, \bibinfo{person}{Christy Doran}, {and} \bibinfo{person}{Thamar Solorio}} (Eds.). \bibinfo{publisher}{Association for Computational Linguistics}, \bibinfo{address}{Minneapolis, Minnesota}, \bibinfo{pages}{119--132}.
\newblock
\href{https://doi.org/10.18653/v1/N19-1011}{doi:\nolinkurl{10.18653/v1/N19-1011}}


\bibitem[Koepke et~al\mbox{.}(2023)]%
        {CE}
\bibfield{author}{\bibinfo{person}{A.~Sophia Koepke}, \bibinfo{person}{Andreea-Maria Oncescu}, \bibinfo{person}{João~F. Henriques}, \bibinfo{person}{Zeynep Akata}, {and} \bibinfo{person}{Samuel Albanie}.} \bibinfo{year}{2023}\natexlab{}.
\newblock \showarticletitle{Audio Retrieval With Natural Language Queries: A Benchmark Study}.
\newblock \bibinfo{journal}{\emph{IEEE Transactions on Multimedia}}  \bibinfo{volume}{25} (\bibinfo{year}{2023}), \bibinfo{pages}{2675--2685}.
\newblock


\bibitem[Kusupati et~al\mbox{.}(2022)]%
        {kusupati2022matryoshka}
\bibfield{author}{\bibinfo{person}{Aditya Kusupati}, \bibinfo{person}{Gantavya Bhatt}, \bibinfo{person}{Aniket Rege}, \bibinfo{person}{Matthew Wallingford}, \bibinfo{person}{Aditya Sinha}, \bibinfo{person}{Vivek Ramanujan}, \bibinfo{person}{William Howard-Snyder}, \bibinfo{person}{Kaifeng Chen}, \bibinfo{person}{Sham Kakade}, \bibinfo{person}{Prateek Jain}, {et~al\mbox{.}}} \bibinfo{year}{2022}\natexlab{}.
\newblock \showarticletitle{Matryoshka Representation Learning}. In \bibinfo{booktitle}{\emph{Advances in Neural Information Processing Systems}}.
\newblock


\bibitem[Kwon et~al\mbox{.}(2023)]%
        {vllm}
\bibfield{author}{\bibinfo{person}{Woosuk Kwon}, \bibinfo{person}{Zhuohan Li}, \bibinfo{person}{Siyuan Zhuang}, \bibinfo{person}{Ying Sheng}, \bibinfo{person}{Lianmin Zheng}, \bibinfo{person}{Cody~Hao Yu}, \bibinfo{person}{Joseph Gonzalez}, \bibinfo{person}{Hao Zhang}, {and} \bibinfo{person}{Ion Stoica}.} \bibinfo{year}{2023}\natexlab{}.
\newblock \showarticletitle{Efficient Memory Management for Large Language Model Serving with PagedAttention}. In \bibinfo{booktitle}{\emph{Proceedings of the 29th Symposium on Operating Systems Principles}} (Koblenz, Germany) \emph{(\bibinfo{series}{SOSP '23})}. \bibinfo{publisher}{Association for Computing Machinery}, \bibinfo{address}{New York, NY, USA}, \bibinfo{pages}{611–626}.
\newblock
\showISBNx{9798400702297}


\bibitem[Lin et~al\mbox{.}(2021)]%
        {pyserini}
\bibfield{author}{\bibinfo{person}{Jimmy Lin}, \bibinfo{person}{Xueguang Ma}, \bibinfo{person}{Sheng-Chieh Lin}, \bibinfo{person}{Jheng-Hong Yang}, \bibinfo{person}{Ronak Pradeep}, {and} \bibinfo{person}{Rodrigo Nogueira}.} \bibinfo{year}{2021}\natexlab{}.
\newblock \showarticletitle{Pyserini: A Python Toolkit for Reproducible Information Retrieval Research with Sparse and Dense Representations}. In \bibinfo{booktitle}{\emph{Proceedings of the 44th International ACM SIGIR Conference on Research and Development in Information Retrieval}} (Virtual Event, Canada) \emph{(\bibinfo{series}{SIGIR '21})}. \bibinfo{publisher}{Association for Computing Machinery}, \bibinfo{address}{New York, NY, USA}, \bibinfo{pages}{2356–2362}.
\newblock
\showISBNx{9781450380379}
\href{https://doi.org/10.1145/3404835.3463238}{doi:\nolinkurl{10.1145/3404835.3463238}}


\bibitem[Lin et~al\mbox{.}(2025)]%
        {lin2025universal}
\bibfield{author}{\bibinfo{person}{Sheng-Chieh Lin}, \bibinfo{person}{Chankyu Lee}, \bibinfo{person}{Mohammad Shoeybi}, \bibinfo{person}{Jimmy Lin}, \bibinfo{person}{Bryan Catanzaro}, {and} \bibinfo{person}{Wei Ping}.} \bibinfo{year}{2025}\natexlab{}.
\newblock \showarticletitle{Universal Multimodal Retrieval with Multimodal {LLM}s}. In \bibinfo{booktitle}{\emph{The Thirteenth International Conference on Learning Representations}}.
\newblock


\bibitem[Ma et~al\mbox{.}(2024a)]%
        {ma-etal-2024-unifying}
\bibfield{author}{\bibinfo{person}{Xueguang Ma}, \bibinfo{person}{Sheng-Chieh Lin}, \bibinfo{person}{Minghan Li}, \bibinfo{person}{Wenhu Chen}, {and} \bibinfo{person}{Jimmy Lin}.} \bibinfo{year}{2024}\natexlab{a}.
\newblock \showarticletitle{Unifying Multimodal Retrieval via Document Screenshot Embedding}. In \bibinfo{booktitle}{\emph{Proceedings of the 2024 Conference on Empirical Methods in Natural Language Processing}}, \bibfield{editor}{\bibinfo{person}{Yaser Al-Onaizan}, \bibinfo{person}{Mohit Bansal}, {and} \bibinfo{person}{Yun-Nung Chen}} (Eds.). \bibinfo{publisher}{Association for Computational Linguistics}, \bibinfo{address}{Miami, Florida, USA}, \bibinfo{pages}{6492--6505}.
\newblock


\bibitem[Ma et~al\mbox{.}(2024b)]%
        {repllama}
\bibfield{author}{\bibinfo{person}{Xueguang Ma}, \bibinfo{person}{Liang Wang}, \bibinfo{person}{Nan Yang}, \bibinfo{person}{Furu Wei}, {and} \bibinfo{person}{Jimmy Lin}.} \bibinfo{year}{2024}\natexlab{b}.
\newblock \showarticletitle{Fine-Tuning LLaMA for Multi-Stage Text Retrieval}. In \bibinfo{booktitle}{\emph{Proceedings of the 47th International ACM SIGIR Conference on Research and Development in Information Retrieval}} (Washington DC, USA) \emph{(\bibinfo{series}{SIGIR '24})}. \bibinfo{publisher}{Association for Computing Machinery}, \bibinfo{address}{New York, NY, USA}, \bibinfo{pages}{2421–2425}.
\newblock
\showISBNx{9798400704314}


\bibitem[Muennighoff et~al\mbox{.}(2023)]%
        {muennighoff-etal-2023-mteb}
\bibfield{author}{\bibinfo{person}{Niklas Muennighoff}, \bibinfo{person}{Nouamane Tazi}, \bibinfo{person}{Loic Magne}, {and} \bibinfo{person}{Nils Reimers}.} \bibinfo{year}{2023}\natexlab{}.
\newblock \showarticletitle{{MTEB}: Massive Text Embedding Benchmark}. In \bibinfo{booktitle}{\emph{Proceedings of the 17th Conference of the European Chapter of the Association for Computational Linguistics}}, \bibfield{editor}{\bibinfo{person}{Andreas Vlachos} {and} \bibinfo{person}{Isabelle Augenstein}} (Eds.). \bibinfo{publisher}{Association for Computational Linguistics}, \bibinfo{address}{Dubrovnik, Croatia}, \bibinfo{pages}{2014--2037}.
\newblock


\bibitem[Portillo-Quintero et~al\mbox{.}(2021)]%
        {clip4video}
\bibfield{author}{\bibinfo{person}{Jes\'{u}s~Andr\'{e}s Portillo-Quintero}, \bibinfo{person}{Jos\'{e}~Carlos Ortiz-Bayliss}, {and} \bibinfo{person}{Hugo Terashima-Mar\'{\i}n}.} \bibinfo{year}{2021}\natexlab{}.
\newblock \showarticletitle{A Straightforward Framework for Video Retrieval Using CLIP}. In \bibinfo{booktitle}{\emph{Pattern Recognition: 13th Mexican Conference, MCPR 2021, Mexico City, Mexico, June 23–26, 2021, Proceedings}} (Mexico City, Mexico). \bibinfo{publisher}{Springer-Verlag}, \bibinfo{address}{Berlin, Heidelberg}, \bibinfo{pages}{3–12}.
\newblock
\showISBNx{978-3-030-77003-7}


\bibitem[Rasley et~al\mbox{.}(2020)]%
        {deepspeed}
\bibfield{author}{\bibinfo{person}{Jeff Rasley}, \bibinfo{person}{Samyam Rajbhandari}, \bibinfo{person}{Olatunji Ruwase}, {and} \bibinfo{person}{Yuxiong He}.} \bibinfo{year}{2020}\natexlab{}.
\newblock \showarticletitle{DeepSpeed: System Optimizations Enable Training Deep Learning Models with Over 100 Billion Parameters}. In \bibinfo{booktitle}{\emph{Proceedings of the 26th ACM SIGKDD International Conference on Knowledge Discovery \& Data Mining}} (Virtual Event, CA, USA) \emph{(\bibinfo{series}{KDD '20})}. \bibinfo{publisher}{Association for Computing Machinery}, \bibinfo{address}{New York, NY, USA}, \bibinfo{pages}{3505–3506}.
\newblock
\showISBNx{9781450379984}


\bibitem[Reimers and Gurevych(2019)]%
        {reimers-2019-sentence-bert}
\bibfield{author}{\bibinfo{person}{Nils Reimers} {and} \bibinfo{person}{Iryna Gurevych}.} \bibinfo{year}{2019}\natexlab{}.
\newblock \showarticletitle{Sentence-BERT: Sentence Embeddings using Siamese BERT-Networks}. In \bibinfo{booktitle}{\emph{Proceedings of the 2019 Conference on Empirical Methods in Natural Language Processing}}. \bibinfo{publisher}{Association for Computational Linguistics}.
\newblock


\bibitem[Sharifymoghaddam et~al\mbox{.}(2025)]%
        {rankllm}
\bibfield{author}{\bibinfo{person}{Sahel Sharifymoghaddam}, \bibinfo{person}{Ronak Pradeep}, \bibinfo{person}{Andre Slavescu}, \bibinfo{person}{Ryan Nguyen}, \bibinfo{person}{Andrew Xu}, \bibinfo{person}{Zijian Chen}, \bibinfo{person}{Yilin Zhang}, \bibinfo{person}{Yidi Chen}, \bibinfo{person}{Jasper Xian}, {and} \bibinfo{person}{Jimmy Lin}.} \bibinfo{year}{2025}\natexlab{}.
\newblock \showarticletitle{{R}ank-{LLM}: A Python Package for Reranking with LLMs}. In \bibinfo{booktitle}{\emph{Proceedings of the 48th International ACM SIGIR Conference on Research and Development in Information Retrieval}} (Pandova, Italy) \emph{(\bibinfo{series}{SIGIR '25})}. \bibinfo{publisher}{Association for Computing Machinery}, \bibinfo{address}{Pandova, Italy}.
\newblock


\bibitem[Sun et~al\mbox{.}(2023)]%
        {sun-etal-2023-chatgpt}
\bibfield{author}{\bibinfo{person}{Weiwei Sun}, \bibinfo{person}{Lingyong Yan}, \bibinfo{person}{Xinyu Ma}, \bibinfo{person}{Shuaiqiang Wang}, \bibinfo{person}{Pengjie Ren}, \bibinfo{person}{Zhumin Chen}, \bibinfo{person}{Dawei Yin}, {and} \bibinfo{person}{Zhaochun Ren}.} \bibinfo{year}{2023}\natexlab{}.
\newblock \showarticletitle{Is {C}hat{GPT} Good at Search? Investigating Large Language Models as Re-Ranking Agents}. In \bibinfo{booktitle}{\emph{Proceedings of the 2023 Conference on Empirical Methods in Natural Language Processing}}, \bibfield{editor}{\bibinfo{person}{Houda Bouamor}, \bibinfo{person}{Juan Pino}, {and} \bibinfo{person}{Kalika Bali}} (Eds.). \bibinfo{publisher}{Association for Computational Linguistics}, \bibinfo{address}{Singapore}, \bibinfo{pages}{14918--14937}.
\newblock


\bibitem[Team(2025)]%
        {qwen2.5-VL}
\bibfield{author}{\bibinfo{person}{Qwen Team}.} \bibinfo{year}{2025}\natexlab{}.
\newblock \bibinfo{title}{Qwen2.5-VL}.
\newblock


\bibitem[Thakur et~al\mbox{.}(2021)]%
        {thakur2021beir}
\bibfield{author}{\bibinfo{person}{Nandan Thakur}, \bibinfo{person}{Nils Reimers}, \bibinfo{person}{Andreas R{\"u}ckl{\'e}}, \bibinfo{person}{Abhishek Srivastava}, {and} \bibinfo{person}{Iryna Gurevych}.} \bibinfo{year}{2021}\natexlab{}.
\newblock \showarticletitle{{BEIR}: A Heterogeneous Benchmark for Zero-shot Evaluation of Information Retrieval Models}. In \bibinfo{booktitle}{\emph{Thirty-fifth Conference on Neural Information Processing Systems Datasets and Benchmarks Track (Round 2)}}.
\newblock


\bibitem[Wang et~al\mbox{.}(2024)]%
        {wang2024improvingtextembeddingslarge}
\bibfield{author}{\bibinfo{person}{Liang Wang}, \bibinfo{person}{Nan Yang}, \bibinfo{person}{Xiaolong Huang}, \bibinfo{person}{Linjun Yang}, \bibinfo{person}{Rangan Majumder}, {and} \bibinfo{person}{Furu Wei}.} \bibinfo{year}{2024}\natexlab{}.
\newblock \showarticletitle{Improving Text Embeddings with Large Language Models}.
\newblock \bibinfo{journal}{\emph{arXiv:2401.00368}} (\bibinfo{year}{2024}).
\newblock


\bibitem[Xu et~al\mbox{.}(2025)]%
        {Qwen2.5-Omni}
\bibfield{author}{\bibinfo{person}{Jin Xu}, \bibinfo{person}{Zhifang Guo}, \bibinfo{person}{Jinzheng He}, \bibinfo{person}{Hangrui Hu}, \bibinfo{person}{Ting He}, \bibinfo{person}{Shuai Bai}, \bibinfo{person}{Keqin Chen}, \bibinfo{person}{Jialin Wang}, \bibinfo{person}{Yang Fan}, \bibinfo{person}{Kai Dang}, \bibinfo{person}{Bin Zhang}, \bibinfo{person}{Xiong Wang}, \bibinfo{person}{Yunfei Chu}, {and} \bibinfo{person}{Junyang Lin}.} \bibinfo{year}{2025}\natexlab{}.
\newblock \showarticletitle{Qwen2.5-{O}mni Technical Report}.
\newblock \bibinfo{journal}{\emph{arXiv:2503.20215}} (\bibinfo{year}{2025}).
\newblock


\bibitem[Xu et~al\mbox{.}(2016)]%
        {msrvtt}
\bibfield{author}{\bibinfo{person}{Jun Xu}, \bibinfo{person}{Tao Mei}, \bibinfo{person}{Ting Yao}, {and} \bibinfo{person}{Yong Rui}.} \bibinfo{year}{2016}\natexlab{}.
\newblock \showarticletitle{MSR-VTT: A Large Video Description Dataset for Bridging Video and Language}. In \bibinfo{booktitle}{\emph{2016 IEEE Conference on Computer Vision and Pattern Recognition (CVPR)}}. \bibinfo{pages}{5288--5296}.
\newblock


\bibitem[Zhang et~al\mbox{.}(2025a)]%
        {zhang2025rankwogpt}
\bibfield{author}{\bibinfo{person}{Crystina Zhang}, \bibinfo{person}{Sebastian Hofst{\"a}tter}, \bibinfo{person}{Patrick Lewis}, \bibinfo{person}{Raphael Tang}, {and} \bibinfo{person}{Jimmy Lin}.} \bibinfo{year}{2025}\natexlab{a}.
\newblock \showarticletitle{Rank-Without-GPT: Building GPT-Independent Listwise Rerankers on Open-Source Large Language Models}. In \bibinfo{booktitle}{\emph{Advances in Information Retrieval}}, \bibfield{editor}{\bibinfo{person}{Claudia Hauff}, \bibinfo{person}{Craig Macdonald}, \bibinfo{person}{Dietmar Jannach}, \bibinfo{person}{Gabriella Kazai}, \bibinfo{person}{Franco~Maria Nardini}, \bibinfo{person}{Fabio Pinelli}, \bibinfo{person}{Fabrizio Silvestri}, {and} \bibinfo{person}{Nicola Tonellotto}} (Eds.). \bibinfo{publisher}{Springer Nature Switzerland}, \bibinfo{address}{Cham}, \bibinfo{pages}{233--247}.
\newblock
\showISBNx{978-3-031-88711-6}


\bibitem[Zhang et~al\mbox{.}(2025b)]%
        {stella}
\bibfield{author}{\bibinfo{person}{Dun Zhang}, \bibinfo{person}{Jiacheng Li}, \bibinfo{person}{Ziyang Zeng}, {and} \bibinfo{person}{Fulong Wang}.} \bibinfo{year}{2025}\natexlab{b}.
\newblock \showarticletitle{Jasper and Stella: distillation of SOTA embedding models}.
\newblock \bibinfo{journal}{\emph{arXiv:2412.19048}} (\bibinfo{year}{2025}).
\newblock


\bibitem[Zhang et~al\mbox{.}(2023)]%
        {MIRACL}
\bibfield{author}{\bibinfo{person}{Xinyu Zhang}, \bibinfo{person}{Nandan Thakur}, \bibinfo{person}{Odunayo Ogundepo}, \bibinfo{person}{Ehsan Kamalloo}, \bibinfo{person}{David Alfonso-Hermelo}, \bibinfo{person}{Xiaoguang Li}, \bibinfo{person}{Qun Liu}, \bibinfo{person}{Mehdi Rezagholizadeh}, {and} \bibinfo{person}{Jimmy Lin}.} \bibinfo{year}{2023}\natexlab{}.
\newblock \showarticletitle{{MIRACL: A Multilingual Retrieval Dataset Covering 18 Diverse Languages}}.
\newblock \bibinfo{journal}{\emph{Transactions of the Association for Computational Linguistics}}  \bibinfo{volume}{11} (\bibinfo{date}{09} \bibinfo{year}{2023}), \bibinfo{pages}{1114--1131}.
\newblock
\showISSN{2307-387X}


\bibitem[Zhang et~al\mbox{.}(2025c)]%
        {gme}
\bibfield{author}{\bibinfo{person}{Xin Zhang}, \bibinfo{person}{Yanzhao Zhang}, \bibinfo{person}{Wen Xie}, \bibinfo{person}{Mingxin Li}, \bibinfo{person}{Ziqi Dai}, \bibinfo{person}{Dingkun Long}, \bibinfo{person}{Pengjun Xie}, \bibinfo{person}{Meishan Zhang}, \bibinfo{person}{Wenjie Li}, {and} \bibinfo{person}{Min Zhang}.} \bibinfo{year}{2025}\natexlab{c}.
\newblock \showarticletitle{{GME}: Improving Universal Multimodal Retrieval by Multimodal LLMs}.
\newblock \bibinfo{journal}{\emph{arXiv:2412.16855}} (\bibinfo{year}{2025}).
\newblock


\end{thebibliography}




\end{document}